\documentclass[onecolumn,sort&compress,numbers]{els-mrw} % For numbered references Style 

\usepackage{amsmath,amssymb,amsfonts,amsthm,makeidx,graphicx}
\usepackage{txfonts}
\usepackage{helvet}

\usepackage[hidelinks]{hyperref}
\usepackage{graphics}
\usepackage{multirow}
\usepackage{multicol}
\usepackage{caption}
\usepackage{subcaption}

\usepackage{mathtools}
\usepackage{slashed} %for slash notation

\newcommand{\bs}[1]{{\boldsymbol{#1}}}
\newcommand{\psibar}{{\bar\psi}}
\newcommand{\what}[1]{{\widehat{#1}}}

\newcommand{\corr}[1]{{\langle #1\rangle}}
\newcommand{\Tr}[1]{{{\rm Tr}\left\{#1\right\}}}
\newcommand{\tr}[1]{{{\rm tr}\left\{#1\right\}}}
\newcommand{\cZ}{{\cal Z}}

\newcommand{\oA}{{\Bar A}}
\newcommand{\oF}{{\Bar F}}
\newcommand{\ooF}{{\Bar{\Bar F}}}
\newcommand{\ooA}{{\Bar{\Bar A}}}

\usepackage{xcolor}

\begin{document}

%%%%%%%%%%%%%%%%%%%%%%%%%%%%%%%%%%%%%%%%%%%%%%%%%%%%%%%%%%%%%%%%
%% the following items are mandatory: 
%% - title
%% - author names
%% - affiliation details
%% - abstract
%% - keywords

%% Precise, concise, and informative description of the focus of this work. 
%% Avoid abbreviations and formulae in the title
\chapter{Thermal field theory and the QCD Equation of State}\label{chap1}

%% All author names and affiliations, and email address for corresponding author
\author[0]{Matteo Bresciani}%

\address[0]{\orgname{Trinity College Dublin}, 
\orgdiv{School of Mathematics and Hamilton Mathematics Institute}, \orgaddress{Dublin, Ireland}}

%\articletag{Chapter Article tagline: update of previous edition, reprint.}

\maketitle

%%%%%%%%%%%%%%%%%%%%%%%%%%%%%%%%%%%%%%%%%%%%%%%%%%%%%%%%%%%%%%%%
%% the following item is mandatory: 
%% 100-150 word summary of the chapter
\begin{abstract}[Abstract]
This Chapter introduces QCD at finite temperature and density. We first present the formulation 
of the thermal theory in the Euclidean path integral formalism. We then describe how the strong
dynamics at high temperature can be inspected through thermal effective field theories. As a
concrete example of thermodynamic quantity, we discuss the Equation of State, which characterises 
the equilibrium properties of the QCD plasma. We finally conclude with an overview of the phase
diagram of strongly interacting matter.
\end{abstract}

%% 5-10 words that embody the key topics in the chapter. What terms would someone put into a search engine if they were looking for a chapter like this?
\begin{keywords}
 	Thermal field theory \sep QCD \sep Equation of State \sep Phase diagram
\end{keywords}

%%%%%%%%%%%%%%%%%%%%%%%%%%%%%%%%%%%%%%%%%%%%%%%%%%%%%%%%%%%%%%%%
%% the following item is mandatory: 
%% List of the key points and topics a reader can expect to learn from this chapter 
\section*{Objectives}
\begin{itemize}
    \item The formulation of QCD at finite temperature and
    density is first presented.
    \item Then, the theoretical tool of the thermal effective theory is introduced to investigate
    QCD at high temperature.
    \item The Equation of State is introduced and its determination, both with perturbative and 
    non-perturbative methods, is reviewed.
    \item The main properties of QCD matter under different regimes of temperature and density
    are described.
\end{itemize}

\section{Introduction}
\label{sec:Introduction}
In the Standard Model, the strong interactions of quarks and gluons are described by Quantum
Chromodynamics (QCD). Among the rich spectrum of phenomena arising from QCD, those
occurring under extreme conditions of temperature and density are of primary relevance for
improving our understanding of Nature, from particle physics to astrophysics.
At very high temperatures, QCD matter takes the form of a plasma of quarks and gluons, which is 
believed to have characterised the Universe in its earliest stages. The opposite regime of cold 
and very dense matter is instead expected to occur in the core of neutron stars. In particle 
physics, heavy-ion collision experiments investigate QCD matter close to the conditions 
where hadrons melt into the plasma phase.

Accurate predictions about QCD thermodynamics are essential for interpreting experimental 
data~\cite{ALICE:2022hor,Gyulassy:2004zy}, and as inputs for the cosmological 
models~\cite{Saikawa:2018rcs,Saikawa:2020swg}. A key theoretical quantity is the Equation of
State, which encodes the temperature dependence of the effective degrees of freedom of the QCD
plasma. As such, it is crucial for charting the QCD phase diagram at different temperatures and
densities. Moreover, the Equation of State directly influenced the expansion of the early Universe
and it might have left a distinctive footprint in cosmological observables like the primordial
gravitational waves~\cite{Saikawa:2018rcs}.

The aim of this Chapter is to introduce the formalism of QCD thermodynamics and summarise the 
current understanding of the behaviour of strongly interacting matter at non-zero temperature and 
density. In Section~\ref{sec:QCD thermodynamics} we set up the theoretical framework of the
thermal field theory. Section~\ref{sec:Thermal effective theory of QCD} presents
the thermal effective theory approach, a commonly used tool to treat 
separately contributions from the different thermal scales that characterise QCD at high
temperature. This is also a convenient framework for perturbative calculations at finite 
temperature. Section~\ref{sec:QCD Equation of State} is dedicated to the Equation of State: 
we review its determination both in perturbation theory and non-perturbatively using lattice QCD,
and discuss the state of the art in the field. Section~\ref{sec:Thermal phases of QCD} provides 
an overview of the present understanding of the QCD phase diagram based on theoretical and
experimental results, and highlights some open questions. Finally, Section~\ref{sec:Summary}
contains some concluding remarks on the topics discussed throughout the Chapter.

\section{QCD thermodynamics}
\label{sec:QCD thermodynamics}
We are interested in the description of a thermal system of interacting quarks and gluons at a
temperature $T$, quark chemical potential $\mu$ and in a three-dimensional box of volume $V$. At
the quantum mechanical level, such system is described by the grand-canonical partition function 
\begin{equation}
    \cZ = \Tr{e^{-\beta(\what{H} - \mu\what{N})}}\,,
    \label{eq:Zfint}
\end{equation}
where $\beta=1/T$ is the inverse temperature
\footnote{In particle physics the unit system is chosen such that $c=\hbar=k_B=1$,
where $c$ is the speed of light, $\hbar$ the reduced Planck constant and $k_B$ the Boltzmann
constant. All dimensionful quantities can be expressed in terms of one unit only, set to be the 
electronvolt (eV). In particular, energy and temperature have the same unit. For reference, a
temperature of $1\,$eV corresponds to about $10^5$ K.}, $\widehat{H}$ is the Hamiltonian
operator and $\what{N}$ is the quark number operator (three times the baryon number).
A positive (negative) chemical potential thus represents a net baryon (anti-baryon) number. The
trace is taken over all the states in the Hilbert space on which the Hamiltonian $\widehat{H}$
acts. The partition function in Eq.~\eqref{eq:Zfint} admits a representation in terms of a path
integral over the gauge and quark fields in Euclidean spacetime,
\begin{equation}
    \cZ = \int DAD\psibar D\psi\,e^{-S_{\rm QCD}[A, \psibar, \psi]} \,,
    \label{eq:Z_QCD}
\end{equation}
where $A_\mu(x)$ is the gauge field, $\psi(x),\psibar(x)$ are the quark and antiquark fields,
the action reads
\begin{equation}
    S_{\rm QCD}[A, \psibar, \psi] = 
    \int_0^\beta dx_0 \int_V d^3\bs{x}\,{\cal L}_{\rm QCD}[A, \psibar, \psi]
    \label{eq:S_QCD}
\end{equation}
and ${\cal L}_{\rm QCD}$ is the QCD Lagrangian.
The finite temperature thus amounts to restricting the integration over Euclidean time to the
interval $0\leq x_0\leq\beta$ in the definition of $S_{\rm QCD}$. Along this compact direction the 
gauge fields satisfy periodic boundary conditions, 
\begin{equation}
    A_\mu(x_0+\beta, \bs{x}) = A_\mu(x_0, \bs{x})\,,
    \label{eq:bc_glue}
\end{equation}
while the fermionic fields satisfy anti-periodic boundary conditions,
\begin{equation}
    \begin{aligned}
        \psi(x_0+\beta, \bs{x}) = -\psi(x_0, \bs{x})\,, \\
        \psibar(x_0+\beta, \bs{x}) = -\psibar(x_0, \bs{x})\,.
    \end{aligned}
    \label{eq:bc_quark}
\end{equation}
Along the spatial directions, which are assumed to be much larger than the compact one, all
the fields satisfy periodic boundary conditions.
The thermodynamic average of an observable $O$ is then given by the expectation value
\begin{equation}
    \corr{O} = \frac{1}{\cZ} \int DAD\psibar D\psi\,O[A, \psibar,\psi]\,
    e^{-S_{\rm QCD}[A, \psibar, \psi]}\,.
    \label{eq:vOv}
\end{equation} 
As for QCD in the vacuum, the path integral in Eq.~\eqref{eq:vOv} can be
rigorously defined through the lattice regularization of the theory, a renormalization procedure
and the continuum limit. We refer to standard textbooks for the details (for instance,
Refs.~\cite{Montvay:1994cy,Rothe:1992nt,Kapusta:2007xjq}), and use here a formal continuum 
notation.

Given the partition function we can now introduce the thermodynamic functions, which describe
the properties of a plasma of quarks and gluons at equilibrium. The pressure is the most 
fundamental one, whose expression in the thermodynamic limit (i.e. infinite-volume limit) reads
\begin{equation}
    p = \lim_{V\to\infty}\frac{T}{V}\ln\cZ\,,
    \label{eq:pressure}
\end{equation}
and represents a counting of the effective degrees of freedom of the thermal system per unit
volume.
The energy density $e$, the entropy density $s$, and the particle density $n$ of the thermal
system are respectively defined as
\begin{equation}
    e = \lim_{V\to\infty}\frac{T^2}{V}\frac{\partial}{\partial T} \ln \cZ\,, \quad
    s = \frac{\partial p}{\partial T}\,, \quad
    n = \frac{\partial p}{\partial \mu}\,.
    \label{eq:esnf}
\end{equation}
All together, the thermodynamic functions satisfy the relation
\begin{equation}
    Ts + \mu n = e + p\,,
\end{equation}
and their dependence on the external parameters $T$ and $\mu$ goes under the name of the 
Equation of State.

\section{Thermal effective theory of QCD}
\label{sec:Thermal effective theory of QCD}
We discuss now a commonly used theoretical framework to study strong interactions at
asymptotically high temperatures: the thermal effective theory approach to 
QCD~\cite{Appelquist:1981vg,Braaten:1995jr}. In this regime three relevant energy scales emerge:
the hard scale, of $O(\pi T)$, the soft scale, of $O(gT)$ where $g$ is the strong coupling, and 
the ultrasoft scale, of $O(g^2T)$. Asymptotic freedom guarantees that the three scales are 
separated for very high temperatures, and the QCD degrees of freedom can be effectively
factorised in a similar manner. Physics at long distances with respect to the compact
direction (or, equivalently, at low energies with respect to the temperature) can be described
by an effective theory whose degrees of freedom are the soft and ultrasoft modes, while the hard
modes contribute through the matching coefficients. This effective theory is called
Electrostatic QCD. Following similar arguments, Electrostatic QCD can be further reduced to an
effective theory for the ultrasoft modes only, called Magnetostatic QCD. Notice that, given the
classification of the degrees of freedom, the thermal effective theories can be defined 
non-perturbatively and are well suited, in principle, for numerical
simulations~\cite{Hietanen:2004ew,Hietanen:2008tv}. The thermal effective theory is also
the standard framework where perturbative calculations at high temperature are carried out. 

\subsection{Electrostatic QCD}
\label{ssec:Electrostatic QCD}
We consider thermal QCD at zero chemical potential and vanishing quark masses. At very high
temperatures asymptotic freedom allows us to treat the theory in the non-interacting case. 
In momentum space, the free gluon propagator in the Feynman gauge is
\begin{equation}
    (D_G)^{ab}_{\mu\nu}(n,\bs p) = \frac{\delta^{ab}\delta_{\mu\nu}}{|{\bs p}|^2+(2\pi nT)^2}\,,
    \label{eq:DG}
\end{equation}
where $a,b$ are colour indices and $n$ is a non-negative integer labelling the Fourier modes
along the compact direction, called Matsubara frequencies. The free (massless) fermion
propagator reads
\begin{equation}
    D_F(n, \bs p) = \frac{-i(2n+1)\pi T \gamma_0 -ip_k\gamma_k}
    {|{\bs p}|^2+ [(2n+1)\pi T]^2}\,,
    \label{eq:DF}
\end{equation}
where $\gamma_0$ and $\gamma_k$, $k=1,2,3$ denote the Dirac matrices and repeated indices are
summed. We now consider physics at large distances compared to the size
$1/T$ of the compact direction, which means small three-momentum compared to the temperature: 
$|\bs{p}|\ll T$. The fermionic modes are suppressed by the non-vanishing term $(2n+1)\pi T$
(whose $n=0$ contribution can be interpreted as a thermal mass), as well as the gauge modes with
$n>0$: these are called the hard modes. The remaining $n=0$ modes of the gauge field are static,
i.e. time-independent, a fact that is known as dimensional reduction. We denote them as
$\oA_\mu(\bs x)$, and they represent the field content of a three-dimensional effective theory 
of QCD holding at energies much lower than $T$, called Electrostatic QCD (EQCD). Dimensional
reduction leads to different gauge transformation rules for electric and magnetic components,
\begin{align}
    \oA'_0(\bs{x}) &= \Omega(\bs{x})\oA_0(\bs{x})\Omega(\bs{x})^\dagger\,,
    \label{eq:gauge_A0} \\
    \oA'_j(\bs{x}) &= \Omega(\bs{x})\oA_j(\bs{x})\Omega(\bs{x})^\dagger 
    -\frac{i}{g_E}\Omega(\bs{x})\partial_j\Omega^\dagger(\bs{x})\,,
    \label{eq:gauge_Aj} \\
    \oF'_{jk}(\bs{x}) &= \Omega(\bs{x})\oF_{jk}(\bs{x})\Omega(\bs{x})^\dagger\,,
    \label{eq:gauge_Fij}
\end{align}
where $g_E$ is the coupling of EQCD, $\oF_{jk}=(ig_E)^{-1}[D_j,D_k]$ is the field strength 
tensor associated to $\oA_j$, $D_j=\partial_j+ig_E\oA_j$ is the covariant derivative, and 
$\Omega(\bs x)\in{\rm SU}(3)$ is a time-independent gauge transformation. As we will see
below, the mass dimension of the coupling $g_E$ is such that $[g_E^2]=1$. With 
the chosen conventions, this implies that $[\oA_0]=[\oA_j]=1/2$. The EQCD Lagrangian includes
all the Lorentz and gauge-invariant operators with mass dimension $\leq 3$,
\begin{equation}
    {\cal L}_{\rm EQCD} = \frac{1}{4}\tr{\oF_{jk}\oF_{jk}}
    + \tr{D_j\oA_0D_j\oA_0}
    + m^2_E\tr{\oA_0^2} + \lambda_A\tr{\oA_0^2}^2 + \dots\,,
    \label{eq:L_EQCD}
\end{equation}
and the dots stand for terms with higher mass dimension and thus suppressed by the proper power
of $1/T$. As a consequence of dimensional reduction, from 
Eqs.~\eqref{eq:gauge_A0}--\eqref{eq:gauge_Fij} it follows that the component $\oA_0$ effectively 
behaves as a scalar field in the adjoint representation of the gauge group and therefore it admits 
a mass $m_E$, called Debye screening mass. We finally write the action of the effective theory as
\begin{equation}
    S_{\rm EQCD} = \int_V d^3{\bs x}\,{\cal L}_{\rm EQCD}\,.
    \label{eq:S_EQCD}
\end{equation}
In order to match EQCD to QCD, the parameters $g_E^2$, $m_E$ and $\lambda_A$ need to be
expressed in terms of the QCD parameters so that, when the same physical quantity is computed
in the two theories, the same result is obtained up to temperature-suppressed corrections due 
to the (infinitely many) higher dimensional operators that have been omitted in the Lagrangian.
Then, given the matching, theoretical predictions for static quantities can be
obtained by studying the three-dimensional effective theory in
place of full QCD~\cite{Hietanen:2008tv}. If the matching is determined in perturbation theory,
then the EQCD parameters are expressed in terms of the strong coupling $g$ and $T$. At leading
perturbative order and for $N_f$ quark flavours they read~\cite{Nadkarni:1988fh,Landsman:1989be}
\begin{equation}
    g_E^2 = g^2T\,, \quad
    m_E^2 = \frac{1}{3}\left(3+\frac{1}{2}N_f\right)g^2 T^2\,, \quad
    \lambda_A = \frac{1}{4\pi^2}g^4 T\,.
\end{equation}

\subsection{Magnetostatic QCD}
\label{ssec:Magnetostatic QCD}
The Debye mass $m_E\sim gT$ and the dimensionful coupling $g_E^2\sim g^2T$ provide the two 
relevant energy scales of EQCD. The former is called soft scale, and it is associated to the 
electric component $\oA_0$ of the gauge field. The latter is called ultrasoft scale, and the
associated degrees of freedom are the magnetic components $\oA_j$ of the gauge field. 
For sufficiently high temperatures, phenomena at energies much lower than $m_E$ can be 
described by an effective theory where the field $\oA_0$ is static and contributes only
through the matching coefficients. This effective theory is called Magnetostatic QCD (MQCD), 
with action
\begin{equation}
    S_{\rm MQCD} = \int_V d^3\bs{x} \left[\frac{1}{4}\tr{\ooF_{jk}\ooF_{jk}} 
    + \dots\right]\,,
    \label{eq:S_MQCD}
\end{equation}
where $\ooF_{jk}$ is the strength tensor for the field $\ooA_j$ in the fundamental 
representation of the gauge group. The dots stand for higher dimensional operators suppressed
by the proper power of $1/m_E$. MQCD is a three-dimensional SU$(3)$ pure gauge theory, with
coupling $g_M^2$. The latter must be expressed in terms of the EQCD parameters so that the two
theories match.  At leading order in perturbation theory, the matching is simply 
$g_M^2 = g_E^2 = g^2T$. MQCD is a confining theory; consequently, even in the weak coupling regime
of QCD at asymptotically high temperatures, the ultrasoft contributions remain of non-perturbative
nature. The only mass scale of the theory is the coupling itself, and therefore 
every dimensionful quantity in MQCD will be proportional to the proper power of $g_M^2$, with a 
non-perturbative coefficient.

\subsection{The QCD pressure}
\label{ssec:The QCD pressure}
As a showcase, we outline in the following how the thermal effective theories can be employed 
to separate the hard, soft and ultrasoft contributions in the QCD pressure, defined in
Eq.~\eqref{eq:pressure}. At high temperatures, the pressure can be written 
as~\cite{Braaten:1995jr}
\begin{equation}
    p(T) = T\left(c_E(T) + \frac{1}{V}\log\cZ_{\rm EQCD}\right)
    \label{eq:p_QCD_EQCD}
\end{equation}
(the thermodynamic limit is understood),
where $\cZ_{\rm EQCD}$ is the partition funciton for EQCD, 
\begin{equation}
    \cZ_{\rm EQCD} = \int D\oA_0 D\oA_j\, e^{-S_{\rm EQCD}}\,,
\end{equation}
and the action $S_{\rm EQCD}$ is given by Eq.~\eqref{eq:S_EQCD}. In addition to the matching of
the coefficients $g_E, m_E, \lambda_A$, we have to match the coefficient $c_E$ as well, that 
can be interpreted as the unit operator omitted in the effective Lagrangian. 
In Eq.~\eqref{eq:p_QCD_EQCD}, $c_E$ represents the contribution to the pressure from the hard
scale, while the logarithm of the EQCD partition function contains the contributions from the
soft and ultrasoft scales. The two latter scales can be separated by introducing MQCD as the
effective theory of EQCD at energy scales much smaller than the soft scale. The pressure of QCD
can thus be written as 
\begin{equation}
    p(T) = T\left( c_E(T) + c_M(T) + \frac{1}{V}\log\cZ_{\rm MQCD}\right)\,,
    \label{eq:p_QCD_EQCD_MQCD}
\end{equation}
where $\cZ_{\rm MQCD}$ is the partition funciton for MQCD, 
\begin{equation}
    \cZ_{\rm EQCD} = \int D\ooA_j\, e^{-S_{\rm MQCD}}\,,
\end{equation}
and the action $S_{\rm MQCD}$ is given by Eq.~\eqref{eq:S_MQCD}. Again, in addition to the 
matching of the coupling $g_M$, the coefficient $c_M$ must be matched as well. In
Eq.~\eqref{eq:p_QCD_EQCD_MQCD}, $c_M$ represents the contribution to the pressure from the soft
scale, while the logarithm of the MQCD partition function contains the contribution from the
ultrasoft scale. The latter must be determined non-perturbatively in MQCD, which is a 
three-dimensional Yang-Mills theory and thus it is much simpler to simulate on the lattice with
respect to full QCD~\cite{Hietanen:2004ew}. Then, given the matching of all the parameters, 
Eq.~\eqref{eq:p_QCD_EQCD_MQCD} yields the pressure of full QCD in the high temperature regime.

The matching of the parameters has been determined in perturbation theory, see
Ref.~\cite{Kajantie:2002wa} for a review. Actually Eq.~\eqref{eq:p_QCD_EQCD_MQCD} is the starting
point for the computation of the pressure in thermal perturbation theory, as we will discuss
later.

\subsection{Scope of the effective theory}
\label{ssec:Scope of the effective theory}
The effective theory approach holds under the assumption that the hard, soft and ultrasoft energy
scales are well separated among each other, so that the relative degrees of freedom can be
effectively factorised. In other words, the hierarchy of scales
\begin{equation}
    g^2 T/\pi \ll gT \ll \pi T
    \label{eq:scales_hierarchy}
\end{equation}
must be respected. The strong coupling $g$ is a function of a renormalization scale $\bar\mu$, 
and the well-known leading order expression for the three-flavour theory is
\begin{equation}
    \frac{1}{g^2(\bar\mu)} = \frac{9}{8\pi^2}\ln(\bar\mu/\Lambda) + \dots\,,
\end{equation}
where $\Lambda$ is the QCD Lambda-parameter. In the $\overline{\rm MS}$ scheme, and for $N_f=3$,
$\Lambda_{\overline{\rm MS}}=341$ MeV~\cite{Bruno:2017gxd}. In the context of thermal QCD, it is
natural to choose the renormalization scale to be proportional to the temperature. The usual choice 
for the scale is the lowest non-vanishing Matsubara frequency of the gluon modes, $\bar\mu=2\pi T$. 
Then, the coupling is a logarithmically decreasing function of $T$, and for sufficiently high 
temperatures the hierarchy~\eqref{eq:scales_hierarchy} will be in place. For reference, the value 
of the running coupling of QCD with three flavours, renormalized in the $\overline{\rm MS}$ 
scheme, is of $O(1)$ at the electroweak scale, of $O(100)$ GeV. This might signal that only in the 
very high temperature regime the inequalities~\eqref{eq:scales_hierarchy} are actually respected.

\section{QCD Equation of State}
\label{sec:QCD Equation of State}
The QCD Equation of State (EoS), introduced in Section~\ref{sec:QCD thermodynamics}, is a 
fundamental theoretical quantity that describes the collective behaviour of a strongly
interacting plasma at equilibrium. It is a key input for the analysis of data produced in 
heavy ion colliders~\cite{ALICE:2022hor,Gyulassy:2004zy}, as well as for understanding the 
thermal expansion of the early Universe~\cite{Saikawa:2018rcs,Saikawa:2020swg}. We will first
discuss the methods that have been developed for the determination of the EoS at vanishing
chemical potential, both perturbatively and non-perturbatively. We will then comment on 
the extension of these results to the case of a non-vanishing chemical potential.

\subsection{Perturbative methods}
\label{ssec:Perturbative methods}
In QCD, asymptotic freedom in principle allows phenomena occurring at very high energy scales to
be described using perturbation theory. In the thermal theory, the temperature is the natural 
energy scale. However, perturbative calculations at finite temperature are plagued by infrared 
divergences~\cite{Linde:1980ts,Gross:1980br} that arise beyond a certain order in the coupling, 
rendering the expansion ill-defined. In the case of the pressure, this breakdown occurs 
at $O(g^6)$. Although the issue remains, through the thermal effective theories (see 
Section~\ref{sec:Thermal effective theory of QCD}) the expansion can be reorganised in such a way 
that the infrared divergences are transparently isolated in the contributions at the ultrasoft
scale. The hard and soft scales instead contribute through the matching coefficients, which are
not sensitive to infrared physics.

The starting point for the perturbative determination of the pressure $p(T)$ is thus the 
decomposition in Eq.~\eqref{eq:p_QCD_EQCD_MQCD}. The details of the calculation can be found in 
the literature (see e.g. Ref.~\cite{Kajantie:2002wa} and references therein), and go beyond the
scope of this Chapter. Nevertheless, from general arguments it is possible to infer characteristic
features of perturbation theory at finite temperature, whose validity extends beyond the specific
case of the calculation of the pressure.
\begin{itemize}
    \item The term $c_E(T)$ represents the contribution from the hard modes. Its determination
    is part of the matching of EQCD to QCD. The only scale involved is $\sim \pi T$, and $c_E$, as
    well as the parameters in the EQCD Lagrangian, can be computed as a perturbative
    series in $g^2$. By dimensional arguments, $c_E\sim (\pi T)^3$ thus we expect contributions at
    all even powers of $g$ starting from $O(g^0)$.
    \item The term $c_M(T)$ represents the contribution from the soft modes. Its determination
    is part of the matching of MQCD to EQCD. The only scale involved is $m_E\sim gT$, and $c_M$,
    as well as the parameters in the MQCD Lagrangian, can be computed as a power series in 
    terms of dimensionless ratios like $g_E^2/m_E$ or $\lambda_A/m_E$. Notice in particular
    that odd powers of $g$ arise in this expansion. By dimensional arguments, $c_M\sim m_E^3$
    thus we expect contributions at all powers of $g$ starting from $O(g^3)$.
    \item The last term in Eq.~\eqref{eq:p_QCD_EQCD_MQCD} represents the contribution from the 
    ultrasoft modes, and as already mentioned it cannot be determined in perturbation theory.
    Nevertheless, since the only relevant scale is $g_M^2\sim g^2T$, we expect that contribution
    to be proportional to $g_M^6$, with a non-perturbative coefficient. Therefore, it will
    enter the expansion starting from $O(g^6)$. 
\end{itemize}
The pressure of hot QCD can be finally written as a power series in $g$~\cite{Kajantie:2002wa},
\begin{equation}
    \frac{p}{T^4} = \frac{8\pi^2}{45}\sum_{k} p_k \left(\frac{g}{2\pi}\right)^k\,,
    \label{eq:pressure_pt}
\end{equation}
where the coefficients $p_k$ are known for $k=0,1,...,5$, while the $p_6$ is partly known. The
latter includes the non-perturbative effects of the ultrasoft scale~\cite{Hietanen:2004ew} and 
a not yet determined perturbative contribution~\cite{Navarrete:2024ruu}. For reference, we report
the coefficients up to one-loop order:
\begin{equation}
    p_0 = 1 + \frac{21}{32}N_f\,, \quad p_1 = 0\,, \quad 
    p_2 = -\frac{15}{4}\left(1+\frac{5}{12}N_f\right)\,.
\end{equation}
The corrections to this result due to non-vanishing quark masses can be obtained from a
perturbative calculation as well, and they have been determined to one-loop 
order~\cite{Laine:2006cp}.

Generally speaking, the validity of this perturbative expansion is tightly related to the regime
of validity of the thermal effective theory, as discussed in 
Subsection~\ref{ssec:Scope of the effective theory}. In practice, the convergence properties
of the series are investigated by looking at the higher order corrections, which can be 
estimated for instance by comparing results at different perturbative orders or by varying the 
renormalization energy scale $\bar\mu$. A representation of the perturbative expansion of the
entropy density (obtained from the pressure using Eq.~\eqref{eq:esnf}) for increasing orders is 
given in the left panel of Figure~\ref{fig:compare_pt_np}. The shadowed bands correspond to a
variation of $\bar\mu$ by the factors $2$ or $1/2$ with respect to the central value 
$\bar\mu=2\pi T$. The poor agreement of the curves up to very high temperatures signals that
higher order effects are relevant, and the convergence is slow at least up to temperatures of
$O(100)$ GeV.

A mitigation of this behaviour has been obtained through a different approach to handle the
infrared problem, called Hard Thermal Loop (HTL) perturbation 
theory~\cite{Andersen:2010ct,Andersen:2011sf}. Building on the effective theory analysis, 
in the HTL framework two mass parameters are introduced, which represent the quark thermal mass
and the Debye screening mass. The HTL perturbation expansion for the pressure has
been computed to three-loop order, which is the last before the unavoidable infrared
divergences manifest themselves. The right panel of Figure~\ref{fig:compare_pt_np} represents
the HTL prediction for the entropy density in the very high temperature regime, for increasing
loop order. The shadowed bands give the dependence of the curves on the variation of $\bar\mu$ by 
the factors $2$ and $1/2$. At the electroweak scale, the agreement between the two highest orders 
is at the level of one combined sigma.

\subsection{Non-perturbative determination}
\label{ssec:Non-perturbative determination}
\begin{figure*}[t]
\begin{center}
\begin{minipage}{0.5\columnwidth}
\centering
\includegraphics[width=0.93\textwidth]{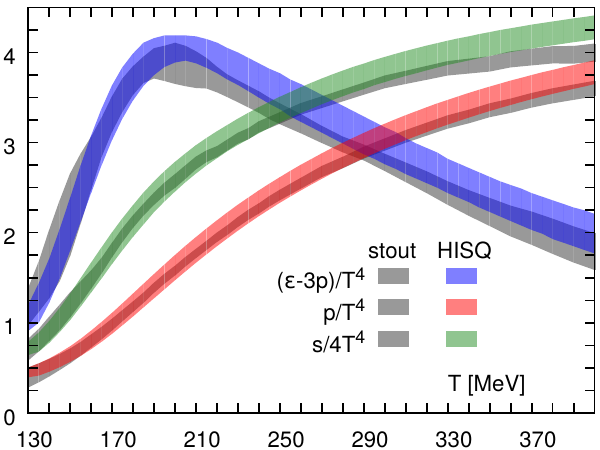}
\end{minipage}%
\begin{minipage}{0.5\columnwidth}
\centering
\vspace{0.2cm}
\includegraphics[width=0.95\textwidth]{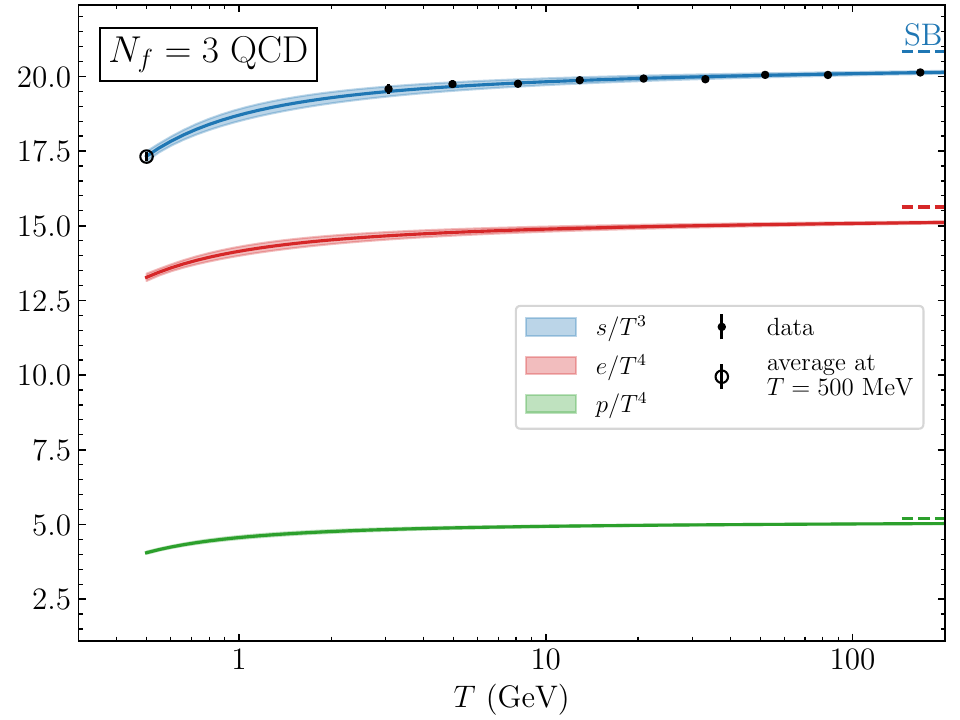}
\end{minipage}
    \caption{Equation of State of QCD with three flavours. Left panel: results for the trace
    anomaly, pressure and entropy density in the interval $130-400$ MeV determined in
    Refs.~\cite{Borsanyi:2013bia,HotQCD:2014kol}. Plot taken from Ref.~\cite{HotQCD:2014kol}. 
    Right panel: results for the entropy density, energy density and pressure for temperatures
    up to the electroweak scale~\cite{Bresciani:2025vxw,Bresciani:2025mcu}. The value at 500 MeV
    is taken from Refs.~\cite{Borsanyi:2013bia,Bazavov:2017dsy}. The dashed lines represent the 
    Stefan-Bolzmann value of the three thermodynamic functions.}
    \label{fig:EoS_Nf3}
\end{center}
\end{figure*}
We discuss now the non-perturbative results for the EoS obtained using lattice QCD.
The pressure is directly related to the partition function of the thermal system, and therefore 
it is defined up to an additive constant to be fixed with some prescription. The usual choice is
that the pressure vanishes at vanishing temperature: $p(T=0)=0$. We can express the pressure
as follows,
\begin{equation}
    \frac{p(T)}{T^4} = \frac{p(T_0)}{T_0^4} + \int_{T_0}^T dT' \frac{I(T')}{(T')^5}\,,
\end{equation}
where $I = e-3p$ is the trace anomaly of the energy-momentum tensor, a quantity suitable for a 
lattice determination by computing expectation values of the pure gauge action and of the
chiral condensate $\corr{\psibar\psi}$. The pressure can thus be obtained by integrating the trace
anomaly, once the latter is known in the continuum limit for several values of the temperature. 
Given pressure and trace anomaly, the other functions follow from standard thermodynamic 
relations. The value $p(T_0)$ of the pressure at the (low) reference temperature $T_0$ can 
be estimated with low-energy models like the hadron resonance gas. This strategy is known
as the integral method, and it was first used for the determination of the 
EoS in the SU($3$) pure gauge theory~\cite{Boyd:1996bx}. In QCD, it has been applied to the
determination of the EoS with $N_f=2+1$ flavours (two light degenerate quarks and a heavier
strange quark)~\cite{Borsanyi:2013bia,HotQCD:2014kol,Bazavov:2017dsy}, and with $N_f=2+1+1$
flavours~\cite{Borsanyi:2016ksw}, up to temperatures of about 1 GeV.

\begin{figure*}[t]
\begin{center}
\begin{minipage}{0.5\columnwidth}
\centering
\includegraphics[width=0.95\textwidth]{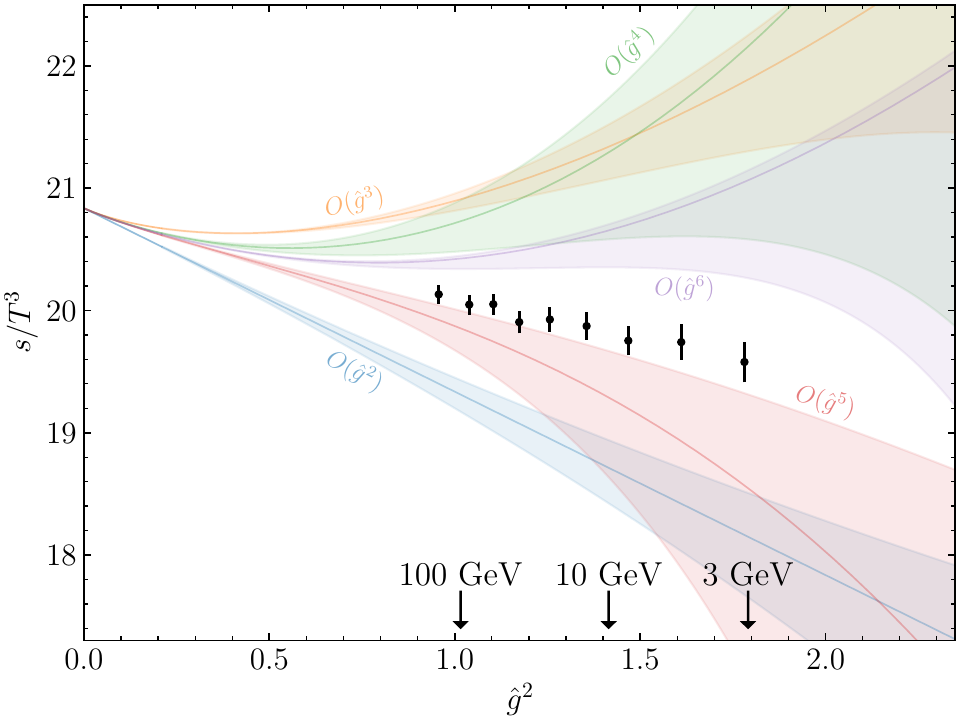}
\end{minipage}%
\begin{minipage}{0.5\columnwidth}
\centering
\includegraphics[width=0.95\textwidth]{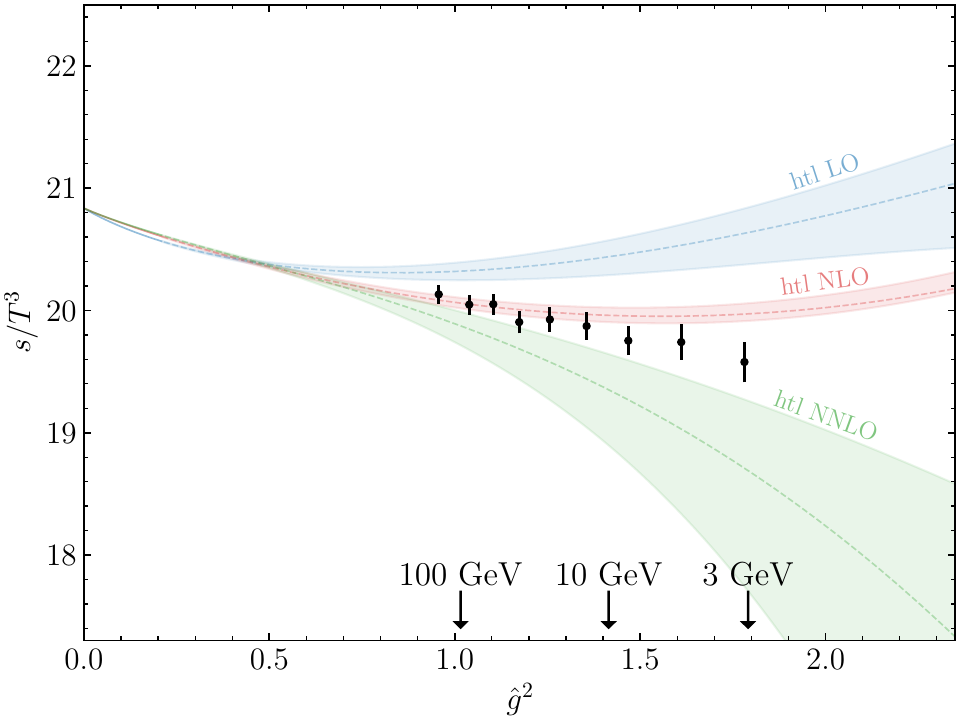}
\end{minipage}
    \caption{Left panel: the non-perturbative results for the entropy density (black dots) from
    Refs.~\cite{Bresciani:2025vxw,Bresciani:2025mcu} are compared to standard perturbation 
    theory. Each curve includes up to the order in the strong coupling (here denoted by 
    $\hat{g}$) reported by the label. 
    Shadowed bands correspond to a variation of the renormalization scale by the 
    factors $2$ or $1/2$. Data are represented as a function of $\hat{g}$, and some values of
    the physical temperature are reported for reference. Right panel: the entropy density is
    compared to HTL perturbation theory at leading order (LO), next-to leading order (NLO) and
    next-to-next-to leading order (NNLO). Shadowed bands have the same meaning as in the left
    panel.}
    \label{fig:compare_pt_np}
\end{center}
\end{figure*}

Even if the integral method is well defined at all temperatures, for technical reasons a precise
lattice calculation following this strategy becomes more and more costly from the computational
point of view as the temperature increases. With the computing resources available to date, in
practice the integral method is limited up to temperatures of the order of 1 GeV.
More recently, a new strategy has been proposed~\cite{Giusti:2016iqr,DallaBrida:2021ddx} 
that overcomes those limitations and allows to explore QCD non-perturbatively for temperatures
up to the electroweak scale. According to this new method, the entropy density can be directly
computed on the lattice by exploiting the Lorentz invariance of the
theory~\cite{Giusti:2010bb,Giusti:2011kt,Giusti:2012yj}. The pressure is then determined by
integrating the entropy in the temperature. Using this strategy the EoS has been determined
for $N_f=3$ QCD up to the electroweak scale~\cite{Bresciani:2025vxw,Bresciani:2025mcu}. 
A summary of the determinations of the EoS with three flavours from different collaborations is
represented in Figure~\ref{fig:EoS_Nf3}. 

The available non-perturbative results at very high temperatures allow us to confront with 
the perturbative expectations in a regime where the latter are supposed to work best.
Figure~\ref{fig:compare_pt_np} compares the results for the entropy density to perturbation
theory in the standard approach (left panel) and in the HTL approach (right panel), as described
in Subsection~\ref{ssec:Perturbative methods}. If we consider both expansions
up to the highest complete order ($O(g^5)$ for the left panel and NNLO for the right
panel), then the agreement with the non-perturbative results is at the level of one to two sigma,
if the statistical error of the data is combined to the systematic one of the expansions due to
the renormalization scale variation. However, a reliable estimation of the latter is difficult
because of the oscillating behaviour of the expansions as the order increases.

\subsection{Finite density}
\label{ssec:Finite density}
In perturbation theory, the non-zero density case can be treated in the limits of large
temperature and small chemical potential, or vanishing temperature and large chemical 
potential~\cite{Freedman:1976ub,Vuorinen:2003fs,Ipp:2006ij,Haque:2013sja}.
Concerning non-perturbative methods, lattice QCD simulations cannot be carried out
straightforwardly at a finite chemical potential, because the latter spoils the probabilistic
interpretation of the Euclidean path integral on which the Monte Carlo sampling is based. This
issue is known as the sign problem, and several strategies have been proposed to obtain results 
at finite density from the lattice at least under the approximation of small ratio $\mu/T$: the
reweighting method~\cite{Ferrenberg:1988yz,Fodor:2001au}, the Taylor expansion
method~\cite{Allton:2002zi}, and the analytic continuation from imaginary chemical
potential~\cite{DElia:2002tig}. For the case of the EoS, the pressure can be Taylor expanded
around $\mu/T=0$, 
\begin{equation}
    \frac{p(T, \mu)}{T^4} = \sum_{n=0}^\infty \hat{p}_{2n}(T) \left(\frac{\mu}{T}\right)^{2n}\,, 
    \quad \hat{p}_n(T)=\frac{1}{n!}\left.\frac{\partial^n (p/T^4)}
    {\partial (\mu/T)^n}\right|_{T,\mu=0}\,,
\end{equation}
where the even powers only contribute because of the reflection symmetry of the partition
function, $\cZ(\mu)=\cZ(-\mu)$. The coefficients $\hat p_n$ are then expressed in terms of 
expectation values of local operators at vanishing chemical potential, that can be determined 
from lattice 
simulations~\cite{Borsanyi:2012cr,Bellwied:2015lba,DElia:2016jqh,Bazavov:2020bjn,Borsanyi:2023wno}.

\section{Thermal phases of QCD}
\label{sec:Thermal phases of QCD}
\begin{figure*}[t]
\begin{center}
\begin{minipage}{0.5\columnwidth}
\centering
\includegraphics[width=0.95\textwidth]{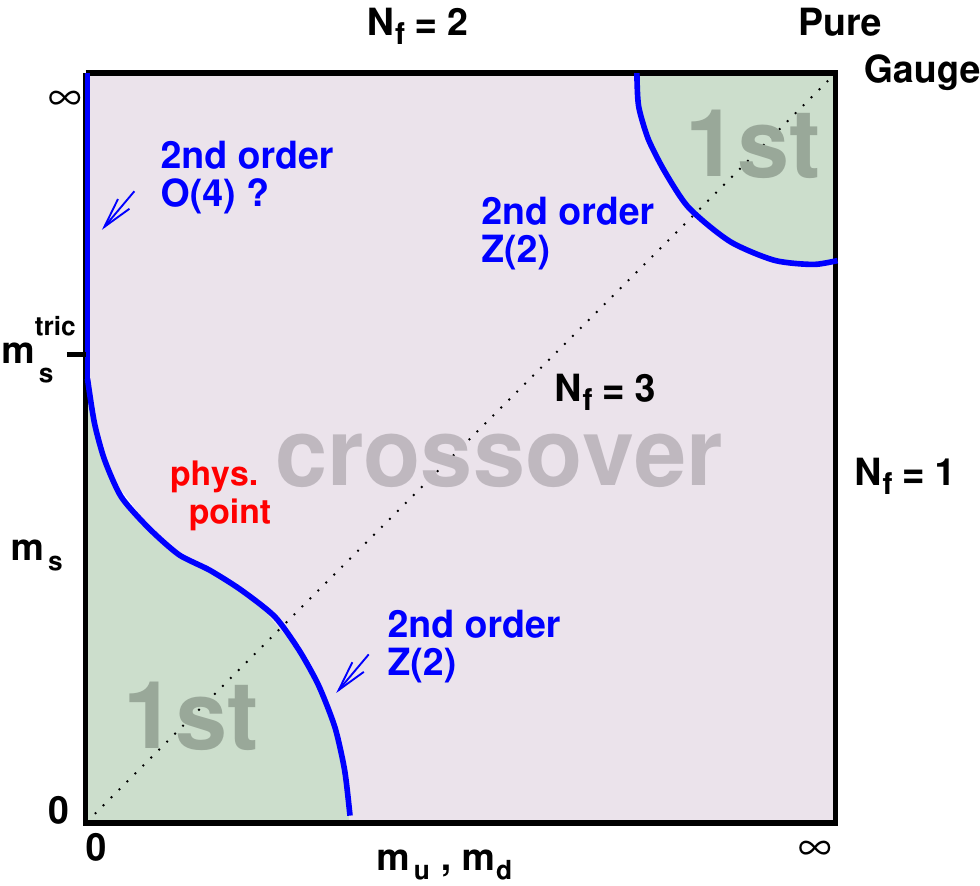}
\end{minipage}%
\begin{minipage}{0.5\columnwidth}
\centering
\includegraphics[width=0.95\textwidth]{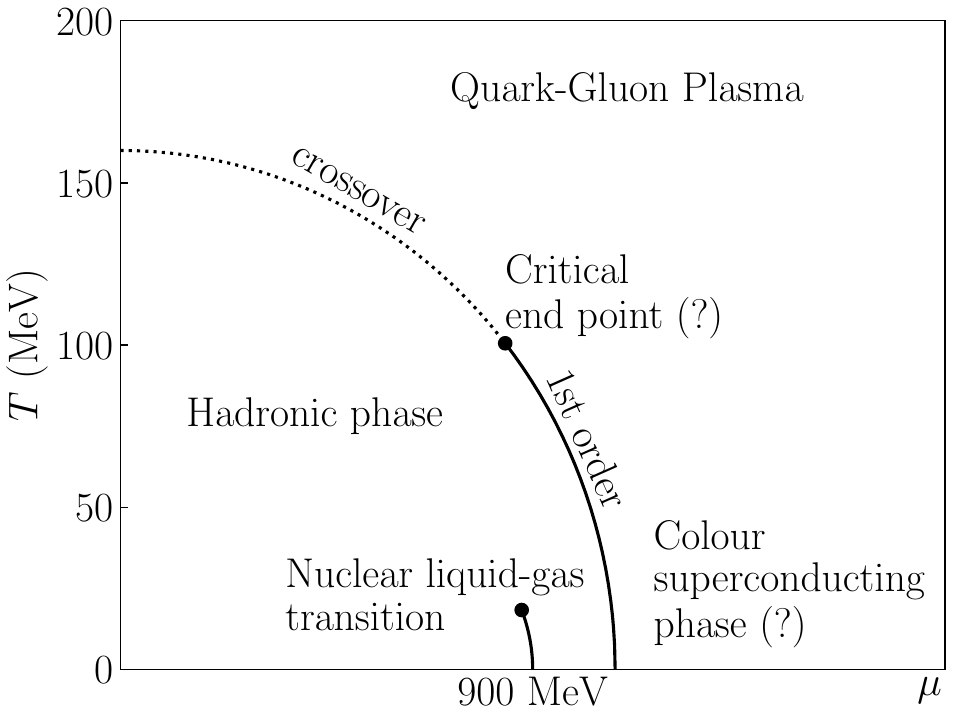}
\end{minipage}
    \caption{Left: Columbia plot from Ref.~\cite{Philipsen:2010gj}. Right: conjectured phase
    diagram of QCD.}
    \label{fig:Columbia_phase}
\end{center}
\end{figure*}
Strongly interacting matter behaves in qualitatively different ways as the temperature and density
change. The various regimes as functions of $T$ and $\mu$ are collected by the QCD 
phase diagram. Unfortunately, at present many properties of the latter remain at the 
conjectural level. On the theoretical side, lattice QCD has proven to be a powerful tool to
explore the strong dynamics non-perturbatively at finite temperature and zero chemical potential.
As discussed in Subsection~\ref{ssec:Finite density}, the sign problem limits the applicability
of lattice QCD at non-vanishing density. Leveraging on asymptotic freedom, partial insights at
very high temperatures or densities may come from perturbative approaches.
On the experimental side, heavy-ion colliders probe the QCD plasma at temperatures up to a few 
hundreds MeV and up to the density of heavy nuclear matter.
The study of compact astrophysical objects like neutron stars may help to chart the QCD 
phase diagram at values of baryon density that cannot be reached by particle physics experiments.

In the following Subsections we overview some established features of the QCD phase diagram,
treating separately the cases of vanishing and non-vanishing chemical potential.

\subsection{The $\mu=0$ case}
\label{ssec:Phase mu=0 case}
There are experimental and theoretical evidences~\cite{Harris:2023tti,Aoki:2006we} that, when
the temperature increases, QCD matter undergoes a crossover from the ordinary hadronic matter 
to the so-called quark-gluon plasma, where the relevant degrees of freedom are strongly 
interacting quarks and gluons. For this reason the two regimes are also known as confined and 
deconfined phase, respectively. The pseudocritical temperature at which the crossover occurs
is $T_{pc}\sim 160$ MeV, and it has been estimated in lattice 
QCD~\cite{HotQCD:2018pds,Borsanyi:2020fev}.

The nature of this change of regime is better understood if we consider its dependence on the quark
masses, with the help of the Columbia plot~\cite{Brown:1990ev,Philipsen:2010gj} in the left panel
of Figure~\ref{fig:Columbia_phase}. On the horizontal axis it appears the mass of the light (up
and down) quarks, taken to be degenerate, while the vertical axis is for the strange quark mass.
At energies of the order of $T_{pc}$, the charm, bottom and top quarks can be considered as static
colour sources and excluded from this analysis. 

Let's start from the top-right corner of Figure~\ref{fig:Columbia_phase}, i.e. QCD with 
infinitely heavy quarks. In this scenario all the quarks decouple and QCD behaves as the SU($3$)
pure gauge theory. The latter is known to have a first order phase transition due to the
spontaneous breaking of the $\mathbb{Z}_3$ symmetry, which is the center of the gauge 
group~\cite{Yaffe:1982qf}. 
The related critical temperature $T_c\sim300$ MeV has been determined by non-perturbative
calculations on the lattice~\cite{Celik:1983nc,Borsanyi:2022xml,Giusti:2025fxu}. 
Although the pure gauge theory is unphysical, the value of $T_c$ provides a useful order of
magnitude estimate.

Dynamical quarks break explicitly the $\mathbb{Z}_3$ symmetry. If we allow for a smooth
dependence of the latent heat on the quark masses, the pure gauge point will be
surrounded by a region of first order transition (light green at the top-right corner). 
As we lower the quark masses, the phase transition becomes weaker and weaker till it vanishes
along a critical line of second order phase transition (the blue line). 
The light grey region of the plot is characterised by the absence of a phase transition. 
The low and high temperature phases are smoothly connected by a crossover at some 
pseudo-critical temperature $T_{pc}$, which is in general a function of the quark masses. 
As we mentioned before, the physical point falls within this regime.

For vanishing quark masses (bottom-left corner) the relevant symmetry is chiral symmetry. In
QCD at the chiral point ($m_u=m_d=m_s=0$) the non-singlet axial generators of chiral symmetry are
spontaneously broken at zero temperature. It has been shown from lattice calculations that this
symmetry is restored as the temperature increases~\cite{HotQCD:2012vvd,Laudicina:2022ghk}. It is
however not clear how the restoration occurs: the prediction of a first order phase transition 
for $N_f\geq 3$~\cite{Pisarski:1983ms} has not been confirmed yet (nor completely ruled out) by
non-perturbative calculations~\cite{Cuteri:2021ikv}.

Finally, the situation is not yet settled also on the left edge of the Columbia plot, where
$m_u=m_d=0$ and $m_s>0$. In that case the nature of the transition depends on the
anomalous ${\rm U}(1)$ axial symmetry and on the non-singlet axial generators of the two-flavour
chiral group~\cite{Butti:2003nu,Pelissetto:2013hqa}. For sufficiently high strange quark mass,
the restoration of the non-singlet axial generators should lead to a second order phase
transition (as reported in Figure~\ref{fig:Columbia_phase}). If however the axial anomaly is
suppressed at the critical temperature, then the transition could be either first or second 
order. Non-perturbative explorations of this issue are ongoing, see 
Refs.~\cite{Cuteri:2021ikv,Brandt:2016daq,HotQCD:2019xnw,Kotov:2021rah} for recent results.

\subsection{The $\mu\neq0$ case}
\label{ssec:Phase mu neq 0}
The non-zero chemical potential introduces in principle one extra axis to the picture
described for $\mu=0$. If we set the light and strange quark masses to their physical values, 
the resulting conjectured phase diagram of QCD is represented, in the $T-\mu$ plane, in the 
right panel of Figure~\ref{fig:Columbia_phase}. A huge effort is devoted to the characterisation
of the (pseudo)critical line that separates the hadronic and plasma phases. The curvature can 
be expressed as a change in the temperature $T_{pc}$ at non-vanishing chemical potential. 
For small $\mu/T$ we can expand as follows, 
\begin{equation}
    \frac{T_{pc}(\mu)}{T_{pc}(0)} = 1-\kappa_2\left(\frac{\mu}{T_{pc}(0)}\right)^2 
    -\kappa_4\left(\frac{\mu}{T_{pc}(0)}\right)^4 + \cdots\,,
    \label{eq:Tpc_mu}
\end{equation}
and the coefficients $\kappa_2,\kappa_4$ have been determined from lattice QCD 
calculations by studying the $\mu$-dependence of observables that are sensitive to the 
change of regime~\cite{Bonati:2018nut,HotQCD:2018pds,Borsanyi:2020fev,Smecca:2024gpu}.
It is believed that the crossover becomes stronger and stronger for increasing chemical 
potential, and that for sufficiently high $\mu$ it will eventually turn into a first order 
phase transition. The two regimes should be linked by a critical point in the phase diagram. 
No clear signs of the existence of the critical point have been observed yet, both from
theory and experiments (recent updates in Refs.~\cite{Borsanyi:2025dyp,STAR:2025zdq}). 
At lower temperatures, the nuclear liquid-gas coexistence line at $\mu\sim 920$ MeV
ends with a critical point at a temperature of about 18 MeV~\cite{Elliott:2013pna}. In the high
density region, the appearance of new exotic states of QCD matter is 
expected~\cite{Schafer:2000et}. Even if some indirect insights may be obtained from the study of
neutron stars~\cite{MUSES:2023hyz}, our knowledge of the phase diagram in that regime is limited
by the fact that strong interactions at very high density cannot be explored by experiments in
terrestrial laboratories.

\section{Summary}
\label{sec:Summary}
This Chapter introduces the foundations of QCD thermodynamics. The starting point is the 
grand-canonical partition function for strongly interacting particles, and its representation in 
the Euclidean path integral formalism. The temperature $T$ enters as a compact time direction
of length $1/T$, and the ordinary vacuum QCD is recovered in the limit $T\to 0$.
The pressure is directly related to the partition function, while taking derivatives with respect
to $T$ and the chemical potential $\mu$ yields other fundamental thermodynamic functions 
collectively known as the Equation of State.

At high temperature, thermal screening effects emerge~\cite{Appelquist:1981vg,Braaten:1995jr}.
QCD degrees of freedom separate into hard, soft and ultrasoft modes, respectively screened at the
distances $\sim (\pi T)^{-1}$, $\sim (gT)^{-1}$ and $\sim (g^2T)^{-1}$. Harder modes can be
integrated out using effective field theories, leaving simpler descriptions at large distances.
This scale separation leads to an effective decomposition of the QCD partition function into
contributions from each thermal scale.

These screening effects are responsible of the infrared divergences that spoil the 
weak coupling expansions in thermal QCD~\cite{Linde:1980ts}. Partial improvement
is obtained by resumming the expansion using effective theory methods~\cite{Kajantie:2002wa},
though the perturbative series still converges poorly even at electroweak temperatures. This
motivates alternative approaches such as the hard thermal loop perturbation 
theory~\cite{Andersen:2010ct,Andersen:2011sf}.

Non‑perturbative lattice QCD remains crucial for the investigation of QCD at all temperatures.
As an example, we have summarised the results obtained for the Equation of 
State~\cite{Borsanyi:2013bia,HotQCD:2014kol,Bazavov:2017dsy,Borsanyi:2016ksw,Bresciani:2025mcu},
but in principle several other properties of the QCD plasma can be explored, like hadronic 
screening lengths or transport coefficients~\cite{Meyer:2011gj}. Extending lattice QCD to non‑zero
chemical potential is non-trivial because of the sign problem, though results for small $\mu/T$
can be obtained.

Despite major progress in understanding QCD thermodynamics, several open questions remain about its 
phase diagram. At low density, increasing temperature induces a smooth crossover from the hadronic
matter to the quark-gluon plasma. At high density, this may become a first order transition,
potentially with a critical point, though no definitive evidence exists. The very high density 
regions, which are inaccessible to colliders or lattice methods, remain largely unexplored. 
The hope is that the remarkable advances achieved in recent years in the observation and
characterisation of compact astrophysical objects~\cite{MUSES:2023hyz}, may provide some
information on the properties of cold and dense QCD matter.

\begin{ack}[Acknowledgement]%
 I wish to thank Michele Pepe for valuable comments on the manuscript of this Chapter.
\end{ack}

%%%%%%%%%%%%%%%%%%%%%%%%%%%%%%%%%%%%%%%%%
%% Mandatory: Bibliography using bibtex 
% \bibliographystyle{Numbered-Style} %% for Numbered Reference Style
\bibliographystyle{elsarticle-num}
\bibliography{main.bbl}

\end{document}